# Pressure induced magnetic and magnetocaloric properties in NiCoMnSb Heusler alloy

**Ajaya K. Nayak and K. G. Suresh**[*]
*Magnetic Materials Laboratory, Department of Physics, Indian Institute of Technology Bombay, Mumbai-400076, India*

**A. K. Nigam**
*Tata Institute of Fundamental Research, Homi Bhabha Road, Mumbai-400005, India*

**A. A. Coelho and S. Gama**
*Instituto de Física 'Gleb Wataghin,' Universidade Estadual de Campinas-UNICAMP, CP 6165, Campinas 13 083 970, SP, Brazil*

The effect of pressure on the magnetic and the magnetocaloric properties around the martensitic transformation temperature in NiCoMnSb Heusler alloy has been studied. The martensitic transition temperature has significantly shifted to higher temperatures with pressure, whereas the trend is opposite with the application of applied magnetic field. The maximum magnetic entropy change around the martensitic transition temperature for $Ni_{45}Co_5Mn_{38}Sb_{12}$ is 41.4 J/kg K at the ambient pressure, whereas it is 33 J/kg K at 8.5 kbar. We find that by adjusting the Co concentration and applying suitable pressure, NiCoMnSb system can be tuned to achieve giant magnetocaloric effect spread over a large temperature span around the room temperature, thereby making it a potential magnetic refrigerant material for applications.

*Corresponding author (email: suresh@phy.iitb.ac.in)

## I. INTRODUCTION

Recently Ni-Mn based ferromagnetic shape memory alloys have drawn much attention due to their multifunctional properties in various fields. Among these alloys, Ni-Mn-Ga has been studied extensively since it shows large magnetic field induced strain.[1,2] The most interesting property of these alloys is that they undergo a first order structural transition from high magnetic austenite phase to a considerably low magnetic martensitic phase, which gives rise to many important properties like shape memory effect, magnetocaloric effect (MCE) and magnetoresistance.[3-8] As many of these alloys show martensitic transitions and consequently large MCE around room temperature, they can be potential candidates as room temperature magnetic refrigerants. Discovery of this new series with giant MCE has enhanced the prospect of 'near room temperature magnetic refrigeration', which has emerged as an energy-efficient and eco-friendly alternative to the commonly used gas-based refrigeration technology. Therefore, these materials are being considered along with Gd and $Gd_5(Si_{1-x}Ge_x)_4$ for room temperature magnetic cooling.[9]

Recently, several studies have been reported on the magneto-structural and magneto-thermal properties of Ni-Mn-Sb Heusler system.[10-13] In general the off-stoichiometric alloys show first order structural transition around room temperature, giving rise to large MCE. This alloy generally shows large inverse (negative) MCE, *i.e.*, the entropy change is positive on applying the magnetic field. The magnetic moment of the compound mainly arises from the Mn atoms. The Mn-Mn exchange interaction is strongly dependent on the Mn-Mn distance, which can be changed by chemical substitutions or by application of hydrostatic pressure. These changes are expected to have a profound influence on the magnetic properties. Recently the effect of pressure on some of the Ni-Mn based systems have been reported.[14-17] Apart from these, there are only a very few pressure studies on the Heusler systems, especially on their MCE behavior. Recently, we have reported large enhancement of MCE with Co substitution in NiMnSb[18]. In this paper, we report the effect of pressure on the magnetic and magnetocaloric properties of $Ni_{50-x}Co_xMn_{38}Sb_{12}$. It was found that the substitution of Co for Ni reduces the martensitic transition temperature. Since we have focused in the temperature region near room temperature, we have studied the pressure effect in the compounds with x=4 and 5, which have the martensitic transition close to room temperature.

## II. EXPERIMENTAL DETAILS

The methods of preparation of polycrystalline samples of $Ni_{50-x}Co_xMn_{38}Sb_{12}$ and their structural characterization are reported elsewhere[18]. The magnetization measurements under various applied pressures (P) have been performed using a superconducting quantum interference device (SQUID) magnetometer (Quantum Design) attached with a Cu–Be clamp type pressure cell with a maximum pressure of 12 kbar.

## III. RESULTS AND DISCUSSION

The powder x-ray diffraction patterns of the samples with x=4 and 5 taken at room temperature show that they possess the austenite (cubic) phase. However, the compound with x=4 showed traces of martensitic phase, as this phase starts just below the room temperature. No such peaks corresponding to the martensitic phase have been seen in the compound with x=5. The variation of the lattice parameters with Co concentration was negligible. The temperature dependence of magnetization with various pressures is shown in Fig. 1 (a) and (b). For each pressure, the measurements were done in the field cooled (FC) and the field heated (FH) modes in a field of 1 kOe. For better clarity the magnetization curves are shown in the martensitic transition region only. With decrease in temperature, the materials undergo the martensitic transition, followed by a sharp decrease in the magnetization. The temperatures corresponding to the onset and the end of this transition, on cooling, are called the martensitic start ($M_S$) and the martensitic finish ($M_F$) temperatures, as shown in Fig. 1(a) and (b).

It is clear from Fig.1 that the martensitic transition temperature ($M_S$ or $M_F$) increases with pressure in both the compounds, which indicates that the martensitic phase becomes more stable with pressure. This may be due to the lower volume of the martensitic phase as compared to that of the austenite phase. It is important to note that the change in electronic structure due to pressure decides the transition temperature, as martensitic transition is related to the formation of Ni 'd' and Sb 'p' hybrid state.[19] The increase in pressure would enhance the hybridization between Ni and Sb, thereby strengthening the Ni-Sb bond. This necessitates more thermal energy for the martensite-austenite transition to occur, leading to an increase in the martensitic transition temperature. The magnetization at both $M_S$ and $M_F$ is found to decrease with pressure. However there is a significant decrease in the magnetization value of the austenite phase at $M_S$, which leads to a reduction in the magnetization step across the martensitic transition. The fact that the magnetization at $M_S$ is found to decrease considerably with pressure, indicates that there is no significant enhancement of the Curie temperature of the austenite phase ($T_C^A$) with pressure. This is in contrast to a recent report which shows that in NiMnIn[14], the magnetization in the austenite phase remains constant with pressure. The low magnetization in the martensitic phase indicates the presence of some antiferromagnetic component, even at ambient pressure. From the dM/dT vs. T plots, it is found that the Curie temperature of the martensitic phase ($T_C^M$) increases marginally with pressure. It is also observed that the magnetization in the martensitic phase around $M_F$ decreases nominally with pressure. The net result is that, in both the alloys, the sharpness of the martensitic transition decreases with pressure, which is expected to reflect in the MCE. However, the thermal hysteresis observed between FC and FH at all pressures indicates that the first order nature of the structural transition from austenite to martensitic phase is retained even at the highest pressure.

As mentioned earlier, the magnetic ordering in these compounds are mainly determined by the Mn-Mn indirect exchange interaction, which depends on the Mn-Mn distance. Because of the off-stoichiometry, a part of the Mn atoms may occupy the Sb sites[20]. At ambient pressure, the exchange interaction is known to be ferromagnetic within the Mn sublattices while the interaction between the Mn atoms occupying the regular Mn sites and Sb sites is antiferromagnetic, in both the austenite and martensitic phases. The net result is the competition between the two types of interactions. Depending on these interactions, the austenite to martensitic transition temperature and the magnetization in both austenite and martensitic phases show different behaviors with substitution of Co (for Ni), application of pressure and field etc.

In order to get an understanding on the effect of magnetic field and pressure on the martensitic transition, the thermo-magnetic variation with different fields, at ambient pressure, is shown in Fig. 2. It can be seen that the martensitic transition shifts to lower temperatures with increase in field. For $Ni_{45}Co_5Mn_{38}Sb_{12}$, $M_S$ is 270 K in a field of 1 kOe, which decreases to 252 K at 80 kOe. Therefore, by comparing figures 1 and 2, it is clear that the effect of applied field is opposite to that of pressure. This fact is also highlighted in Fig 3, which shows the variation of $M_S$ with pressure, whereas the inset shows the shift of $M_S$ with field. The difference between the application of field and pressure is also reflected in the magnetization jump ($\Delta M$) between $M_S$ and $M_F$. While the $\Delta M$ increases with field, it decreases with pressure, in both the compounds. These observations indicate that the pressure stabilizes the martensitic state, whereas the magnetic field prefers the austenite state. The increase in the martensitic transition temperature ($\Delta T$) with pressure in the case of x=4 is about 25 K, while it is 28 K for x=5, for the highest pressure. On the other hand, the decrease is only about 15 K even at the highest field of 80 kOe at ambient pressure in x=5. It is also found that the shift in $M_S$ with pressure in the present case is much larger than the one reported in other Ni-Mn based Heusler alloys.

Fig. 4 shows the magnetization isotherms at the ambient (P=0) and the highest pressures for $Ni_{50-x}Co_xMn_{38}Sb_{12}$ with x=4 and 5. For x=4, at P=0 (Fig. 4a) the magnetization isotherms at 292-294 K shows a metamagnetic transition, which is absent at P=9 kbar. In the case of x=5, the metamagnetic transition seen at ambient pressure is present even at higher pressures, as is evident from Fig. 4c-d. The metamagnetic character in the *M-H* isotherms is associated with a field-induced reverse martensitic transformation from a low magnetization martensitic state to a higher magnetization austenite state. In general, it can be seen that the critical fields for the metamagnetic transition increases with pressure, which implies that the strength of the antiferromagnetic component in the martensitic phase increases with pressure. Decrease in the Mn-Mn bond length must be the reason for this increase. Another feature seen from Fig. 4 is that the magnetization value at the highest field as well as the difference in the values between two consecutive isotherms around the martensitic transition temperature decrease with pressure. This reduction would determine the MCE of these compounds with pressure.

The magnetocaloric effect ($\Delta S_M$) was calculated from the isothermal magnetization curves, using the Maxwell equation

$$\Delta S_M(T,H) = \int_0^H \left(\frac{\partial M(T,H)}{\partial T}\right)_H dH \qquad (1)$$

Fig 5(a) and (b) show the $\Delta S_M$ as a function of temperature for applied fields ($\Delta H$) of 20 and 50 kOe for $Ni_{46}Co_4Mn_{38}Sb_{12}$ and $Ni_{45}Co_5Mn_{38}Sb_{12}$ respectively. It can be seen that the entropy change is positive and that the $(\Delta S_M)_{max}$ decreases with pressure. For $Ni_{46}Co_4Mn_{38}Sb_{12}$ there is a continuous decrease in $(\Delta S_M)_{max}$ from 32.3 J/kg K for P=0 to 16.5 J/kg K for P=9 kbar. For $Ni_{45}Co_5Mn_{38}Sb_{12}$ at P=0, the $(\Delta S_M)_{max}$ is 41.4 J/kg K which increases to 46 J/kg K at p=1.1 kbar and finally decreases to 32.5 J/kg K at P=8.5 kbar. The decrease in the MCE can be attributed to the weakening of the first order transition, which results in smaller difference in the magnetization between the austenite and the martensitic phases. This causes a reduction in the $\partial M / \partial T$ value. Another important observation from Fig. 5 is the tunability of the martensitic transition temperature with pressure, even while retaining significant entropy change. This result in large MCE over a wide temperature spans around the room temperature and makes this system quite interesting from the point of view of applications

It can also be seen from Fig. 5 that, at any pressure, the entropy change is larger in x=5, as compared to that in x=4. This may be due to the fact that Co has a larger magnetic moment as compared to Ni and hence substitution of Co for Ni would increase the net magnetic moment. As mentioned earlier, there exist both FM and AFM interactions between the Mn atoms at different sublattices. Substitution of Co for Ni would increase the ferromagnetic coupling. It has been suggested that the enhancement of ferromagnetic coupling may originate from the altered electronic structure due to the presence of the Co 3*d* electrons.[7]

The suitability of a magnetic refrigerant is judged on the basis of its refrigeration capacity (RC), which can be calculated using the equation

$$\text{RC} = \int_{T_1}^{T_2} (\Delta S_M(T))_{\Delta H} \, dT \qquad (2)$$

where $T_1$ and $T_2$ are the temperatures of the cold and the hot sinks. In the present case, the RC is calculated by integrating $\Delta S_M(T)$ curve over the temperature span corresponding to the full with at half maximum (FWHM) points. It is observed that for both x= 4 and 5, RC decreases marginally with pressure. The maximum RC values obtained at ambient pressure for x=4 and 5 are 95 J/kg and 143 J/kg respectively, which decreases to 78 J/kg and 128 J/kg respectively at highest pressure. We have also calculated the hysteresis loss at ambient pressure for $Ni_{46}Co_4Mn_{38}Sb_{12}$. The average hysteresis loss is estimated to be 21 J/kg for the same temperature interval used for calculating the RC. The large hysteresis loss arises due to the metamagnetic transition in the MH isotherms (Fig. 4). The effective RC value for this compound turns out to be 84 J/kg at ambient pressure. It is found that the RC values observed in $Ni_{50-x}Co_xMn_{38}Sb_{12}$ system is much larger than the values reported for the parent system namely $NiMnSb$[10]. However the RC value in present case cannot be compared with that of $Gd_5(Si_{1-x}Ge_x)_4$[9], as it is calculated in very narrow temperature interval.

## IV. CONCLUSIONS

We have studied the effect of pressure on the magnetic and magnetocaloric properties of NiCoMnSb Heusler alloys. Pressure enhances the stability of the martensitic phase, leading to an upward shift in the martensitic transition temperature. The isothermal magnetic entropy change is found to decrease with pressure, which is a result of the reduction in the magnetization step at the martensitic transition. A detailed neutron diffraction study would be essential to unravel the magnetic moment variation brought about by Co substitution and pressure.

**Figure Caption**

FIG. 1 Temperature dependence of magnetization with various pressures for (a) $Ni_{46}Co_4Mn_{38}Sb_{12}$ (b) $Ni_{45}Co_5Mn_{38}Sb_{12}$.

FIG. 2. Temperature dependence of magnetization with various fields for $Ni_{45}Co_5Mn_{38}Sb_{12}$ at ambient pressure.

FIG. 3 Variation of $M_S$ with pressure for $Ni_{50-x}Co_xMn_{38}Sb_{12}$ (x=4, 5). The inset shows the change in $M_S$ with magnetic field at ambient pressure for x=5.

FIG. 4 Magnetization isotherms of $Ni_{50-x}Co_xMn_{38}Sb_{12}$ for x=4 and 5 at the ambient (P=0) and the highest pressures.

FIG. 5 Magnetic entropy change ($\Delta S_M$) at various pressures as a function of temperature for (a) $Ni_{46}Co_4Mn_{38}Sb_{12}$ and (b) $Ni_{45}Co_5Mn_{38}Sb_{12}$

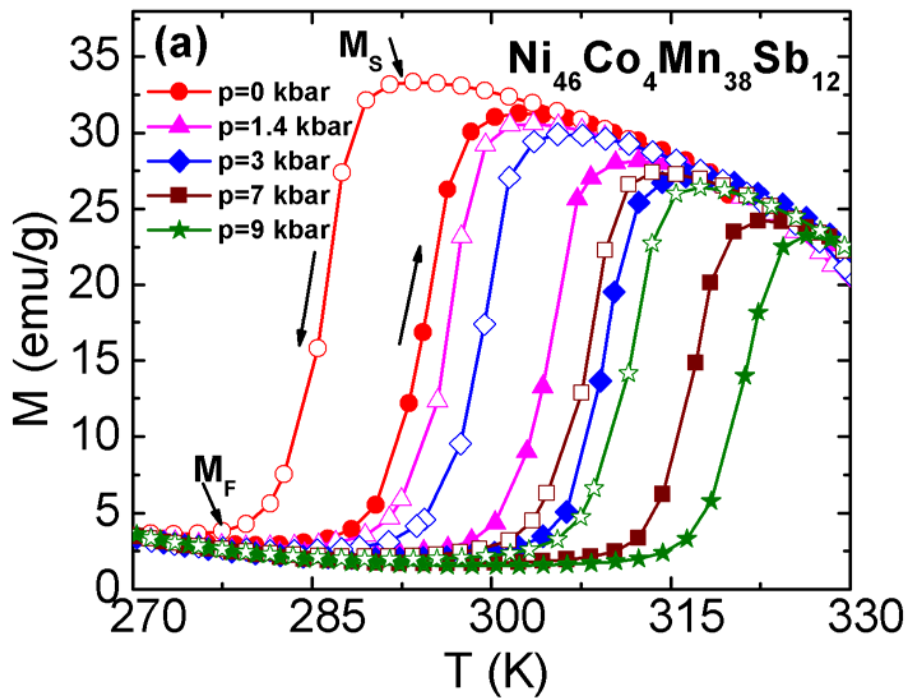

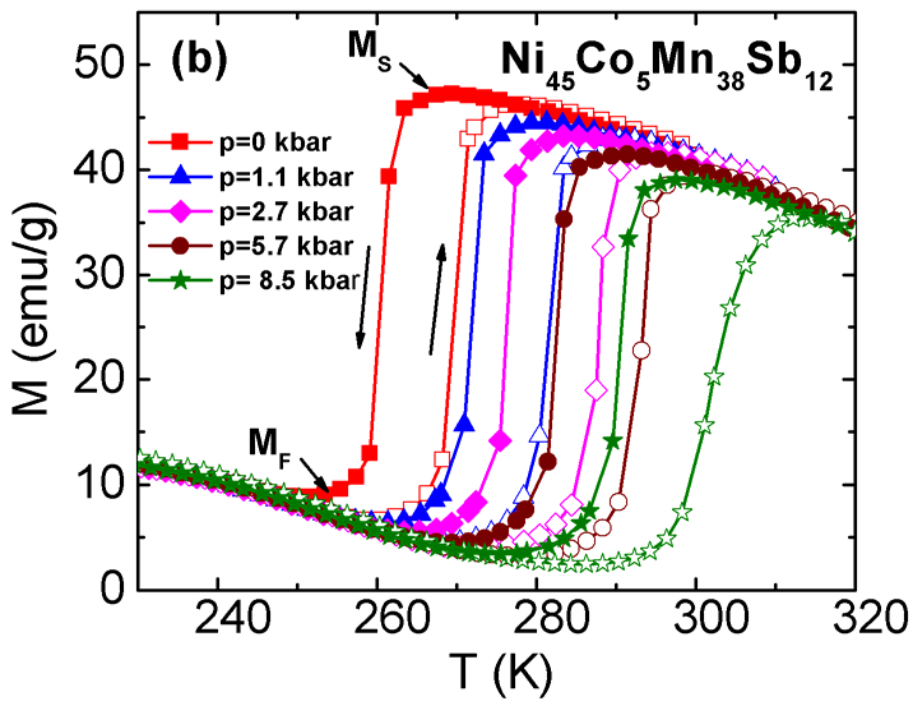

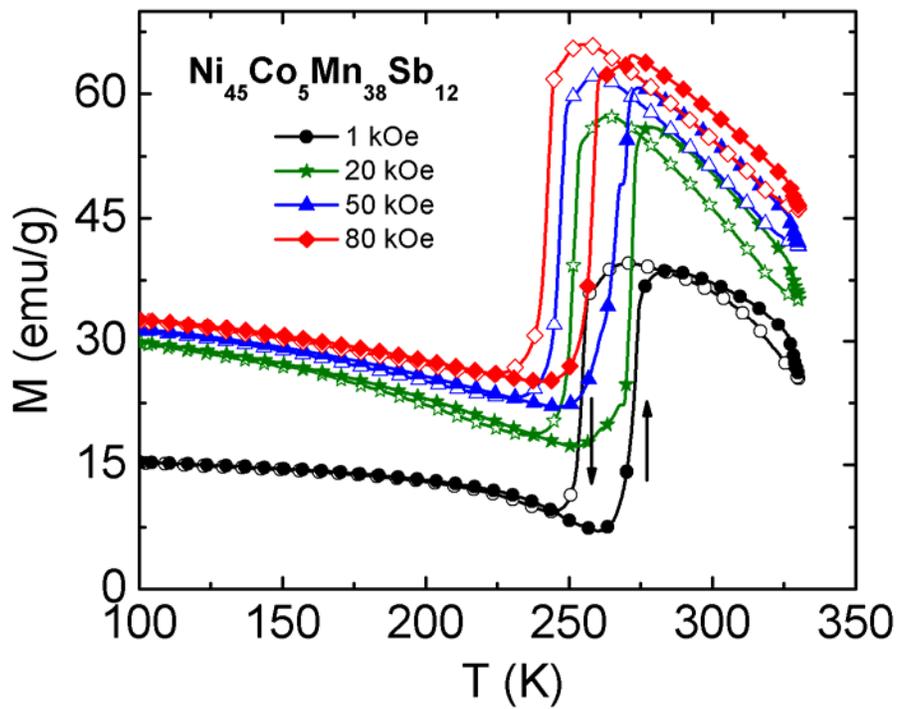

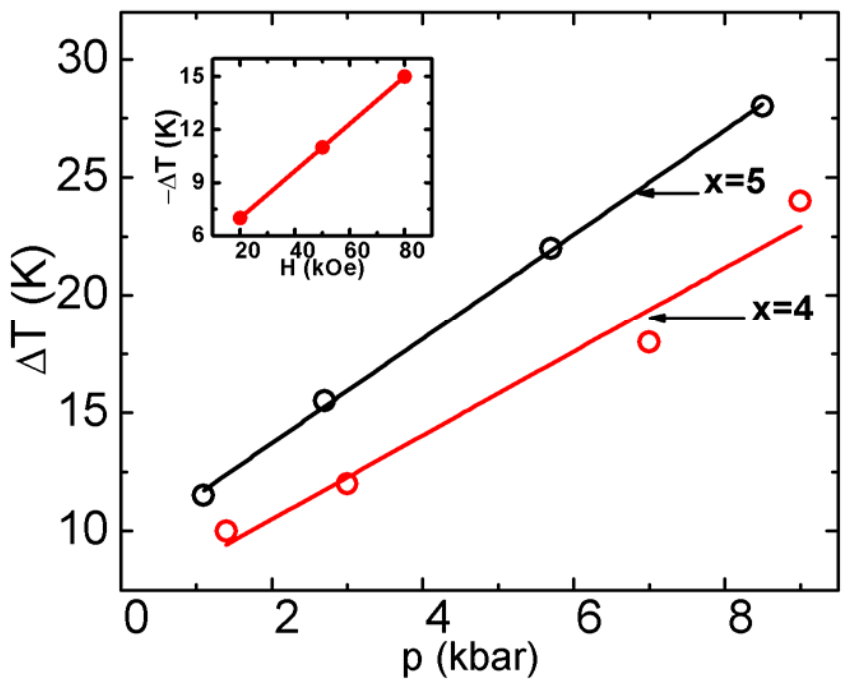

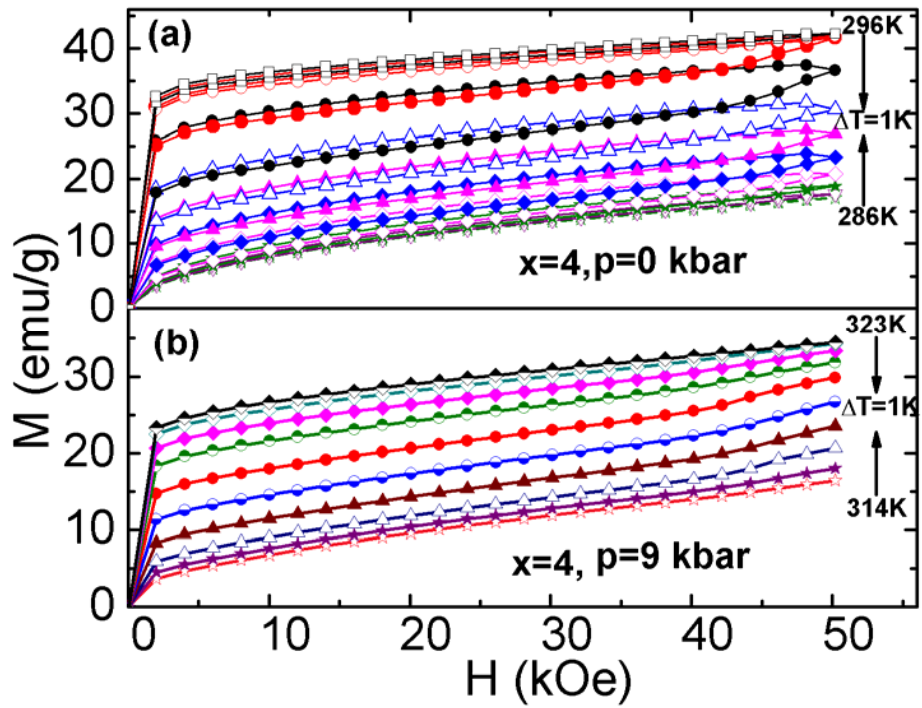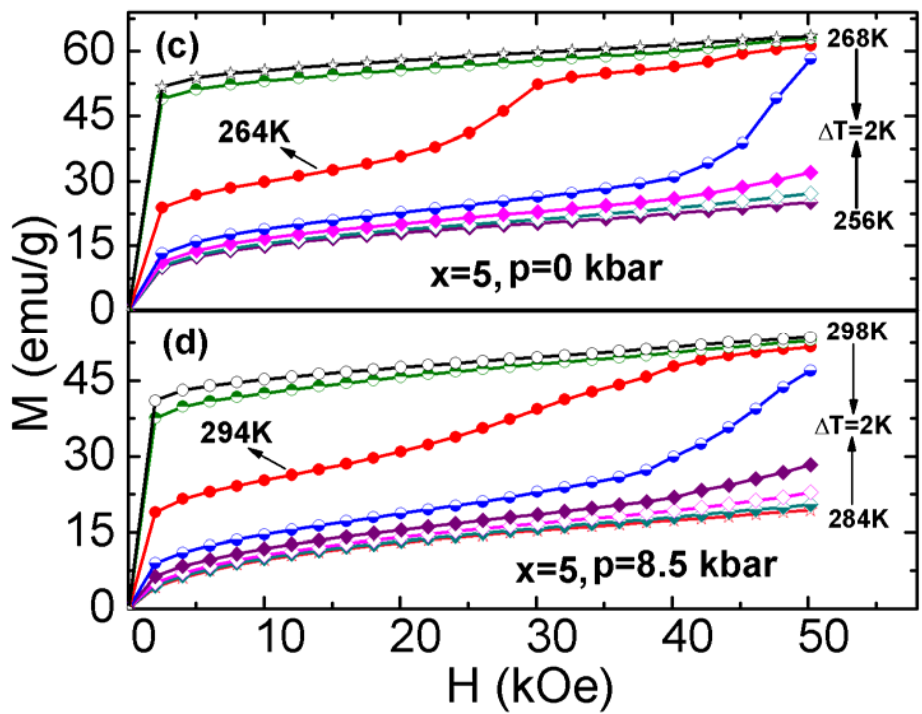

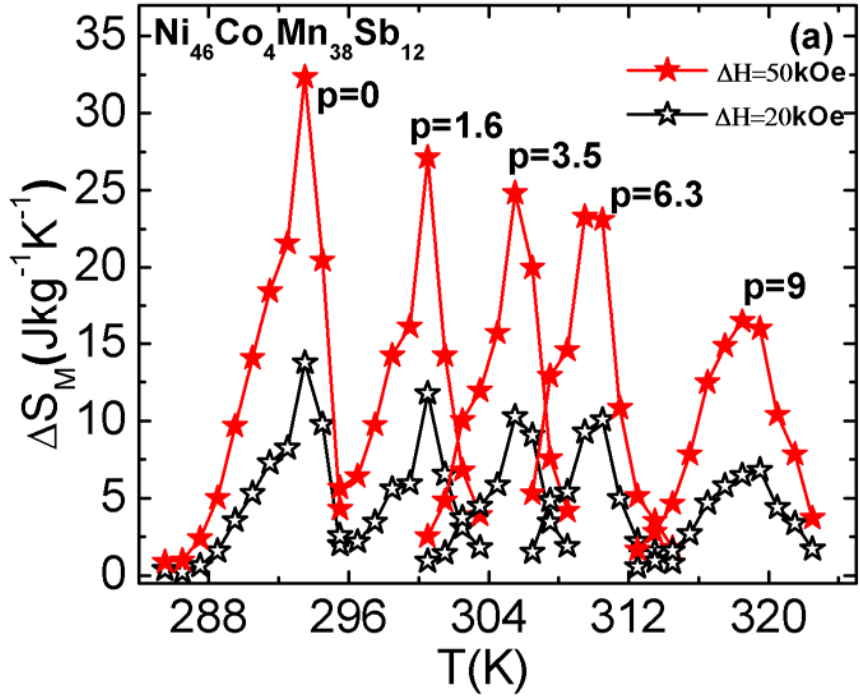

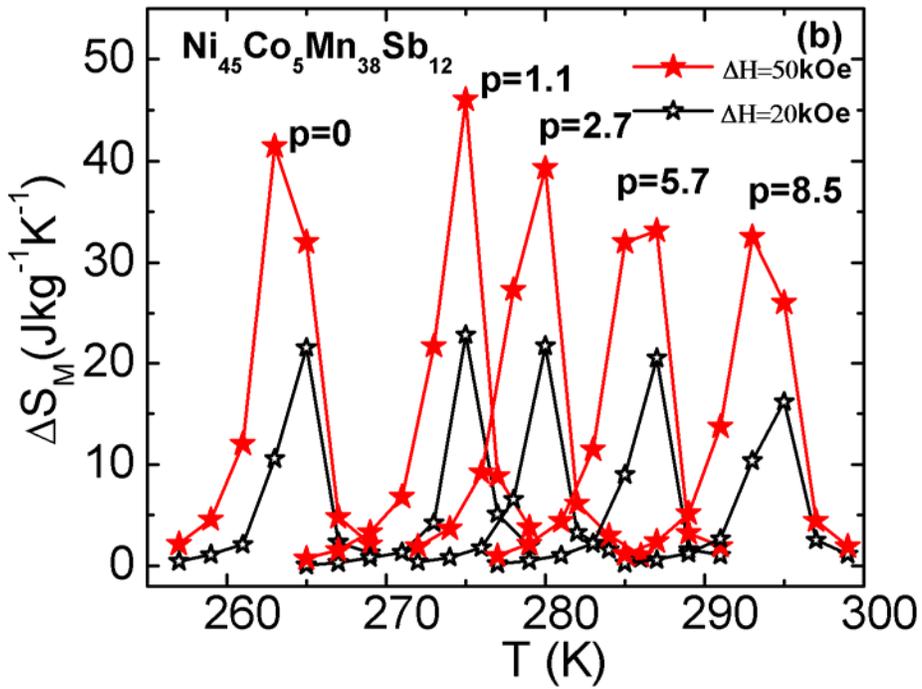